# Elimination, reversal, and directional bias of optical diffraction

Ofer Firstenberg, Paz London, Moshe Shuker, Amiram Ron,

Department of Physics, Technion-Israel Institute of Technology, Haifa 32000, Israel.

Nir Davidson

Department of Physics of Complex Systems, Weizmann Institute of Science, Rehovot 76100, Israel.

## Abstract

We experimentally demonstrate the manipulation of optical diffraction, utilizing the atomic thermal motion in a hot vapor medium of electromagnetically-induced transparency (EIT). By properly tuning the EIT parameters, the refraction induced by the atomic motion may completely counterbalance the paraxial free-space diffraction and by that eliminates the effect of diffraction for arbitrary images. By further manipulation, the diffraction can be doubled, biased asymmetrically to induced deflection, or even reversed. The latter allows an experimental implementation of an analogy to a negative-index lens.

Any image, imprinted on a wave field and propagating in free space, undergoes a paraxial diffraction spreading and eventually blurs. In many disciplines, the possibility to reduce or manipulate the diffraction is explored, for purposes such as imaging, wave guiding, microlithography, and all-optical light processing. As was recently demonstrated, arbitrary images can be imprinted on light pulses which are dramatically slowed when traversing a medium of room-temperature atoms [1,2], via the process of electromagnetically induced transparency [3,4]. In addition to the regular free-space diffraction, the slow-light images undergo diffusion due to the thermal atomic motion [5,6]. Here we report an experimental demonstration of a novel technique to eliminate the paraxial free-space diffraction and the diffusion of slow-light, regardless of its position and shape [7]. By properly tuning the light-matter interaction, the diffraction can be increased, reduced, eliminated completely, or even reversed. Former suggestions for diffraction manipulation of certain modes in vapor have utilized spatial inhomogeneity [8-11] or non-linearity [12]. In contrast, the scheme presented here is linear and occurs only in the wave-vector space, rendering elimination of diffraction for arbitrary images all throughout their propagation. The interaction may be inverted, to accelerate the diffraction in the medium, or biased, to inflict asymmetric diffraction and deflection. Doubling the strength of the interaction surpasses the regular diffraction and effectively reverses it, allowing an implementation of a negative-diffraction lens [13]. Alongside recent advances in slow-light amplification [14] and entanglement of slow images [15], the ability to control diffraction opens various possibilities for classical and quantum image manipulation.

Electromagnetically induced transparency (EIT) is a coherent two-photon interaction between light and atoms [16]. It involves an atomic medium and two light fields, usually a strong 'pump' and a weak 'probe', resonantly coupling two of the atomic levels to a common excited level. When the Raman resonance condition is satisfied, namely when the two-photon frequency-detuning, $\Delta$, is within the atomic spectral-width, $\Gamma$, the atoms are driven towards a 'dark' state, which substantially reduces the absorption in the medium. A short probe pulse propagates in the medium with a reduced group-velocity, owing to the steep dispersion inside the narrow transparency window [4,17-21]. A travelling atomic-coherence field accompanies the probe, and the combined light-matter excitation is termed a dark-state polariton [22]. If an image is imprinted on the probe field in the plane normal to the propagation direction, the complex amplitude of the dark-state polariton follows the amplitude and phase of the image. Free-space diffraction, being essentially a geometric effect, occurs for slow images precisely as it would in free space.

In an EIT medium of hot vapor, the thermal motion of the atoms affects the propagation of images. The addition of a buffer gas attenuates the thermal motion, which becomes diffusive and can be characterized by a single diffusion coefficient, $D$.

As a result, on EIT resonance, the atomic part of the dark-state polariton undergoes diffusion while propagating slowly in the medium. The polariton thus experiences both diffraction and diffusion. Here, we exploit the atomic motion to influence the diffraction. Altering the Raman detuning provides control of the polariton's coupling with atoms moving at a desired direction. We therefore counterbalance diffraction by 'Doppler trapping' the outwards-confronting light components with inwards-moving atoms or, alternatively, force diffraction in a preferable direction.

The free-space diffraction of an optical field envelope $E(x, y; z)$ travelling in the $z$ direction is described in the transverse Fourier plane (Fig. 1a) as $\partial E(\mathbf{k}_\perp; z)/\partial z = [-ik^2/(2q)]E(\mathbf{k}_\perp; z)$, where $k = |\mathbf{k}_\perp|$ is the transverse wave-number, $q = 2\pi/\lambda$, and $\lambda$ the optical wave length. In a vapor EIT medium, the propagation of a weak probe depends on its angular deviation from the pump due to the Doppler-Dicke effect [23,24]. For a wide homogenous non-diverging pump, finite-pump effects such as transverse intensity variation [25] and Ramsey narrowing [26] are made negligible, and the dynamics of the probe becomes [5] $\partial E/\partial z = i\chi(\mathbf{k}_\perp)E$, where $\chi(\mathbf{k}_\perp) = \chi_{\text{EIT}}(\mathbf{k}_\perp) - k^2/(2q)$ is the linear susceptibility and

$$\chi_{\text{EIT}}(\mathbf{k}_\perp) = i\alpha\left(1 - \frac{\Gamma_p}{\Gamma + Dk^2 - i\Delta}\right). \tag{1}$$

Here, $2\alpha$ is the absorption outside the EIT window, $\Gamma_p$ is the power-broadening width, proportional to the pump intensity, and $Dk^2$ is the Doppler-Dicke width. The latter has a simple physical interpretation: Both the residual Doppler broadening and the Dicke narrowing are linear in $k$, which corresponds to the angle between the probe and pump, resulting in a combined quadratic effect [24]. At a given $\Delta$, different $k$-components of a probe image experience different EIT spectra, and consequently the image is altered. When $\Delta=0$ (red lines in Fig. 1b), $\chi_{\text{EIT}}$ is purely imaginary and induces a low-pass absorption filter in $k$ space, with a half width of $k_0 = \sqrt{\Gamma/D}$. For $k \ll k_0$, the filter is quadratic in $k$ and corresponds to standard diffusion, accompanied by the free-space diffraction. In order to manipulate diffraction, a non-zero Raman detuning should be introduced. The case $\Delta=\pm\Gamma$ is of special importance: in this case, the leading quadratic term becomes purely real, inducing diffraction without diffusion,

$$\chi_{\text{EIT}}(\mathbf{k}_\perp) = \chi_{\text{EIT}}(0) \mp \frac{Dk^2}{2v_g} + O(k^4) \quad \text{for } \Delta = \pm\Gamma, \quad (2)$$

with $v_g=\Gamma^2/(\alpha\Gamma_p)$ being the slow group-velocity. Here, the induced diffraction is continuous, in contrast to the diffraction manipulation explored in periodic systems [27-29]. When $\Delta=-\Gamma$, the induced diffraction negates the free-space diffraction, and if $v_g=Dq$, they are cancelled altogether. As demonstrated in Fig. 1b (black lines), for $k\ll k_0$, the susceptibility curves are flat and both diffraction and diffusion are eliminated. In contrast, when $\Delta=+\Gamma$, the induced and the free-space diffraction sum up, increasing the overall diffraction. Note that for given $D$ and $q$, the amount of diffraction is determined exclusively by the group velocity, which is easily controlled by the pump power and the atomic density. A major difficulty of the scheme is the substantial absorption at $\Delta\neq 0$. Fortunately, this absorption is uniform both in real and $k$ spaces and may be compensated for by any linear gain mechanism [7,14].

The experimental setup is depicted in Fig. 1c. A glass vapor cell of length $L$=50 mm is air-heated to 72°C and filled with $^{87}$Rb and 10 Torr of Neon buffer-gas. The rubidium diffusion coefficient is $D$=1100 mm$^2$/s, satisfying the diffraction-elimination condition with $v_g = Dq \cong 8700$ m/s. For our optical depth (~6), we arrive at these conditions with a pump intensity of 660 $\mu$W/mm$^2$, yielding an EIT line-width of $\Gamma$=70 KHz and transmission at $\Delta=\pm\Gamma$ of about 1%. In each shot, an image is imprinted on a weak probe pulse (2 $\mu$W/mm$^2$) and projected onto the entrance facet of the cell. The transmitted probe at the exit facet is recorded with a CCD camera.

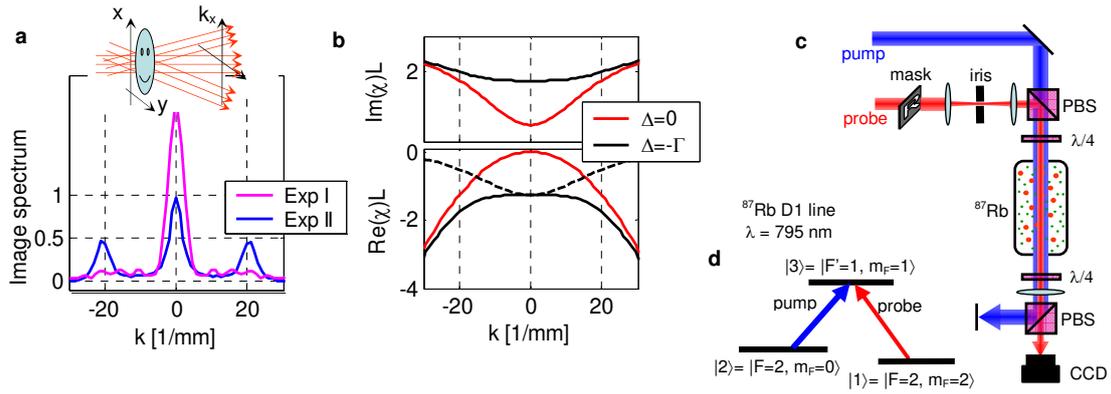

**Figure 1. EIT susceptibility in *k* space and the experimental set-up. a**, An image in the transverse plane, decomposed into its transverse-momentum components. The spectra as a function of *k* correspond to the experiments presented in (I) Fig. 2a and (II) Fig. 2b. **b**, The total susceptibility (solid lines): Imaginary part (top), corresponding to diffusion, and real part (bottom), corresponding to diffraction. For $\Delta=-\Gamma$, both curves are flat up to the fourth order in *k*. The dashed line shows the contribution from the EIT for $\Delta=-\Gamma$, imposing negative diffraction at $k\ll k_0$ and cancelling the free-space diffraction ($k_0=20$ mm$^{-1}$). **c**, The experimental set-up. An image is imprinted on the probe beam using a binary mask, and is imaged onto the entrance facet of the cell by a 4f imaging system. An iris in the central focus point acts as a low-pass filter for the image. The collimated pump and the probe, of orthogonal linear polarizations, are combined on a polarizing beam splitter (PBS), rotated to circular polarizations with a λ/4 plate, passed through the 50-mm long $^{87}$Rb cell, rotated back to linear polarizations and split using a second PBS. The probe is imaged from the exit facet of the cell onto a CCD camera. **d**, The Λ-type level diagram depicts the part of the $^{87}$Rb D1-line that was used in the experiment.

A demonstration of image propagation without diffraction is presented in Fig. 2a with an image of the symbol '®'. The 100-μm features in the image are significantly distorted after 50 mm of free-space diffraction. On the EIT resonance ($\Delta=0$), the image covers this distance in 5.75 μs, during which it diffracts and diffuses. However for $\Delta=-\Gamma$, both diffraction and diffusion are clearly suppressed. The corresponding calculations verify that the observed minor spreading for $\Delta=-\Gamma$ is due to sub-diffractive [29] and sub-diffusive terms of fourth-order in *k*. Indeed, the *k* spectrum of the image extends beyond the $k\ll k_0$ region ($k_0=20$ mm$^{-1}$), as seen in Fig. 1 (Exp I). For $\Delta=+\Gamma$ and as predicted in Eq. (2), the image does not diffuse but rather undergoes substantial diffraction, of effectively twice the distance travelled. The presented calculations are done numerically by taking the Fourier transform of the two-dimensional incident field, multiplying by $\exp[i\chi(\mathbf{k}_\perp)L]$, and taking the inverse Fourier transform. This computation does not require the paraxial $k \ll k_0$ approximation and is hereafter denoted as the numerical calculation.

A slow image of a line grating, shown in Fig. 2b, provides a quantitative measurement of the actual diffraction and diffusion. This image has the property that it reappears periodically after propagating a distance of $L_T = 2a^2/\lambda$, known as the Talbot self-imaging distance, where $a$ is the grating period. At $L_T/2$, a reciprocal grating is created; the original lines disappear and new lines appear in the originally dark areas. In our experiment, the cell length was $\sim L_T/4$, at which the amplitudes of the original and reciprocal gratings are equal, resulting in a grating with a period of $a/2$.

The results for $\Delta = -\Gamma, 0, +\Gamma$ are presented in Fig. 2b. Evidently, the image does not change for $\Delta = -\Gamma$, and it is essential to note that, in contrast to a self-imaging effect, the image is maintained all throughout the propagation. In the $\Delta = 0$ image, diffusion broadens the lines and erases the grating. For $\Delta = +\Gamma$, we observe double diffraction, with the reciprocal grating dominating the original, in correspondence to $L_T/2$. To quantify the amount of diffraction, we measure the intensity at the position of the original and reciprocal gratings, $\langle I_o \rangle$ and $\langle I_n \rangle$ respectively, and define the contrast $C = (\langle I_o \rangle - \langle I_n \rangle)/(\langle I_o \rangle + \langle I_n \rangle)$, shown in Fig. 2b. $C=1$ (–1) at integer (half-integer) Talbot distances, and $C=0$ when the original and reciprocal gratings are comparable. The maximum value of $C$ is obtained at $\Delta = -\Gamma$ and is about 0.85.

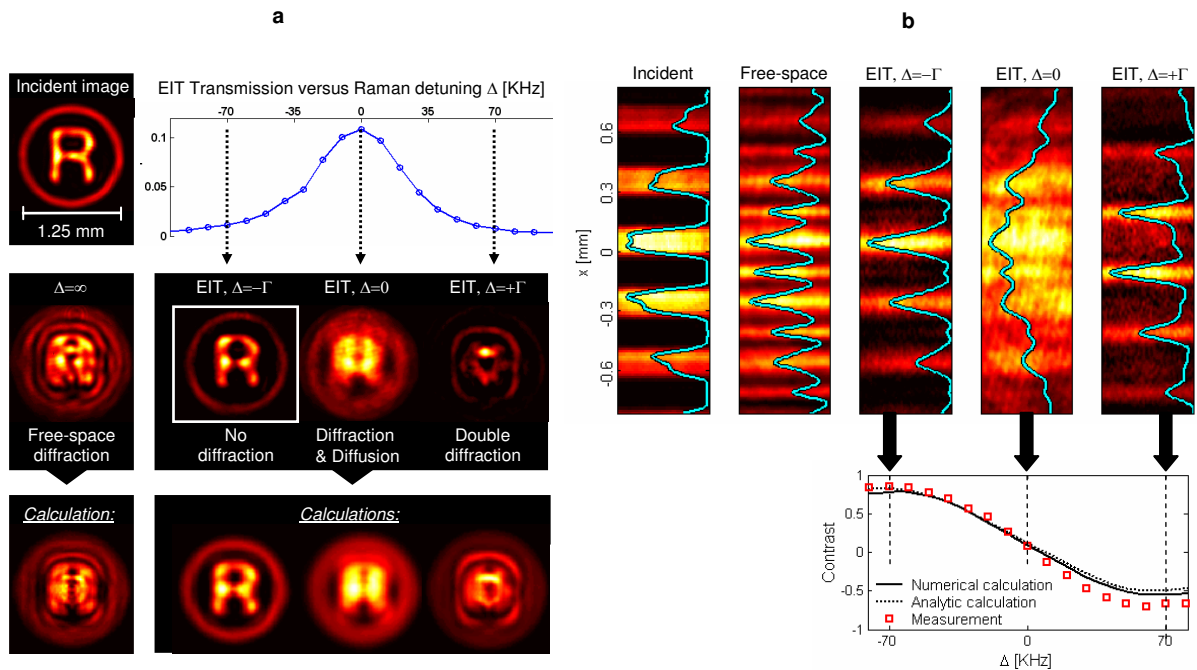

**Figure 2. Elimination of diffraction of arbitrary images. a**, Upper left, an image of the symbol '®' at the entrance facet of the vapor cell. Middle row, the image at the exit facet of the 50-mm cell without EIT (left, taken at off-resonance) and with EIT (right), for $\Delta = -\Gamma$, 0, and $+\Gamma$ ($\Gamma = 70$ KHz). The $\Delta=0$ image exhibits both the regular free-space diffraction and the diffusion associated with the atomic motion. The $\Delta = -\Gamma$ image manifests the elimination of diffraction, as well as the absence of diffusion. The $\Delta = +\Gamma$ image exhibits no diffusion and twice the effect of free-space diffraction – the sum of the free-space and the EIT-induced diffraction. The graph above indicates the transmitted power as a function of $\Delta$. Bottom row, numerical calculations of the effect for each case, given the initial condition of the incident image. **b**, The incident image is an array of lines (3.3 lp/mm), such that after 50 mm of free-space propagation, new lines appear in the dark centers between the original lines (a reciprocal grating at ~1/4 Talbot distance, with $L_T=230$ mm). The cyan curves are transverse cross sections. The elimination of diffraction for $\Delta = -\Gamma$ and the doubling of the diffraction for $\Delta = +\Gamma$ is clearly seen. In the $\Delta = +\Gamma$ image, the original grating almost vanishes, which is equivalent to a 1/2 Talbot distance. In the $\Delta=0$ image, regular free-space diffraction occurs (1/4 Talbot distance), accompanied by diffusion which completely blurs the lines. The bottom graph shows the contrast of the original grating compared to the reciprocal grating, with +1 corresponding to the initial image and -1 to the existence of only the reciprocal grating (1/2 Talbot distance). The contrast is well explained by an analytic analysis, which considers the *k*-spectrum of the image with a single carrier and two sidebands at $\pm 2\pi/a$ (the exact spectrum is shown in Fig. 1a, exp II).

In addition to the magnitude of diffraction, one can utilize the atomic-motion mechanism to control the diffraction directionality. By setting a small angle $\theta_{\text{pump}}$ between the pump and the $z$ axis, as illustrated in Fig. 3a, we superimpose a transverse phase grating, $\exp(ixq\theta_{\text{pump}})$, on the pump-probe interference [5], replacing $Dk^2$ in Eq. (1) with $D(\mathbf{k}_\perp - q\theta_{\text{pump}}\hat{\mathbf{x}})^2$. For $\Delta=\pm\Gamma$ and $\theta_{\text{pump}} \ll (k_0/q)$, the dispersion in Eq. (2) is added a term proportional to $\pm \mathbf{k}_\perp \cdot \hat{\mathbf{x}} \theta_{\text{pump}}$, which inflicts a directional deflection on the probe, in an angle $\theta_{\text{probe}} = \mp (qD/v_g)\theta_{\text{pump}}$. Similarly to the walk-off phenomenon in birefringence crystals, the beam is deflected while the carrier wave-vector stays parallel to the $z$ axis (equal-phase surfaces maintain their original orientation). Hence, the beam keeps its original direction upon exiting the cell.

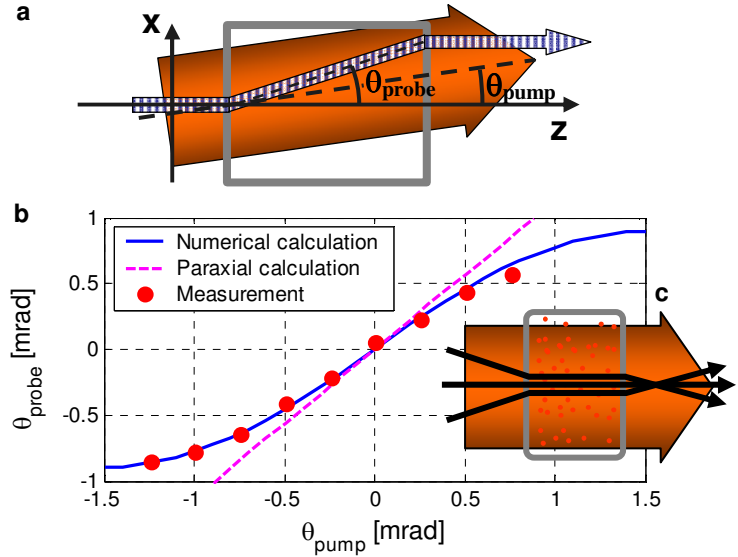

**Figure 3. Induced deflection (walk-off) of the probe beam. a**, The pump is set at an angle $\theta_{\text{pump}}$ relatively to the $z$ axis, and the angular deflection of the probe, $\theta_{\text{probe}}$, is measured. The drawing illustrates the directions for $\Delta<0$. For $\Delta>0$, the probe deflects to the opposite direction. **b**, At the conditions for diffraction elimination ($\Delta=-\Gamma$, $v_g=Dq$) the probe follows the direction of the pump for $q\theta_{\text{pump}} \ll k_0$. The experimental error in determining $\theta_{\text{probe}}$ is ±0.02 mrad; the error in $\theta_{\text{pump}}$ is much smaller. **c**, An explanation of the diffraction elimination via the deflection effect. At these specific conditions, the probe beam 'finds' the pump and refracts in its direction, regardless of the probe's original direction. Thus, all the transverse momentum components of an image refract into the same direction and traverse the cell together, maintaining their phase relation. Upon exiting the cell, each component returns to its original direction.

We have performed deflection experiments in the conditions for diffraction elimination, $v_g = Dq$ and $\Delta = -\Gamma$. Fig. 3b presents the measured $\theta_{probe}$ versus the applied $\theta_{pump}$, showing that $\theta_{probe} = \theta_{pump}$ as long as the paraxial approximation holds ($q\theta_{pump} \ll k_0$). This striking phenomenon, that the probe takes the direction of the pump while in the cell regardless of the incident angle, provides another explanation for the elimination of diffraction. As illustrated in Fig. 3c, all diverging $k$ components of a focused beam are made to propagate in the axial (pump) direction while maintaining their phase relation, thus detaining the diffractive divergence.

Diffraction vanishes when the induced diffraction, $D/(2v_g)$, is equal in size and opposite in sign to the free-space diffraction, $1/(2q)$. However, if the induced diffraction is strengthened, *e.g.*, by lowering the group velocity, the overall diffraction will become negative. Such a medium reverses the diffraction of slow images and can undo diffraction that has already taken place. As spatial diffraction is concerned, a medium with negative diffraction behaves similar to a negative refraction-index medium, within the limits of the paraxial approximation. The effective index is $n_{eff} = (1 - qD/v_g)^{-1}$. When $v_g = qD/2$, $n_{eff} = -1$ and the overall diffraction becomes exactly the opposite of free-space diffraction.

A fascinating application of negative-index materials is an unusual lens [13], demonstrated in Fig. 4. This slab-shaped lens of length $L$ and $n_{eff} = -1$ focuses the radiation coming from any point source, located at a distance $u<L$, to a distance $v$ behind the lens, where $u+v=L$. Parallel light rays, however, are not focused and continue in their original direction. Indeed, the images in the experiment are reconstructed by the lens regardless of the distance $u$.

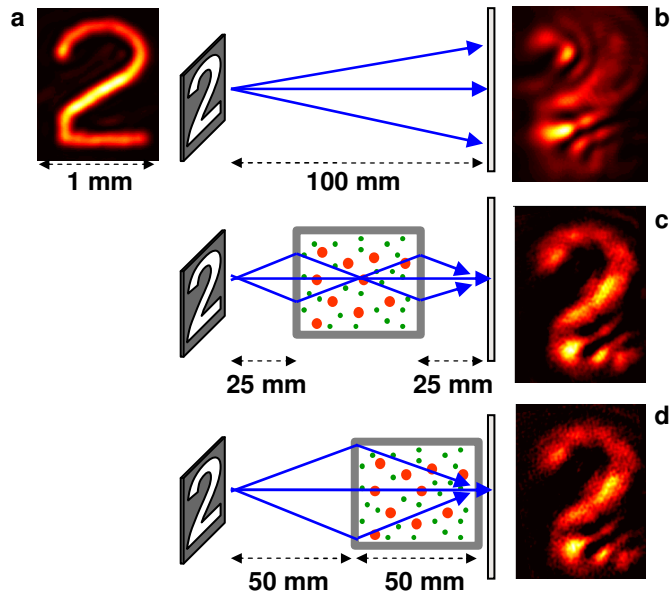

**Figure 4. A negative-diffraction lens.** An image of the digit '2' (**a**) is created at $u$=25 mm in front of the entrance facet of the vapor cell, of length $L$=50 mm, while the CCD is positioned to image the plane that is $v$=25 mm behind the exit facet. The image is significantly blurred under free-space diffraction of 100 mm (**b**). In the negative-diffraction conditions, the 50-mm EIT cell acts as a lens, and the image is made to reappear (**c** and **d**). To achieve these conditions, the pump power is set to about half the power required for the elimination of diffraction, so that $\Gamma$=30 KHz, the group delay is doubled, and $n_{eff} \cong -1$. Because the EIT line-width is decreased, the $k$ filter is narrower ($k_0$=13/mm), making our paraxial approximation less valid and the lens marginally adequate for this purpose. The imperfections in the reconstruction are due to the small $k_0$. When entering the cell, the beams refract in an angle opposite to the incident angle, and refract back upon exiting, in similarity to a negative refraction-index lens. However, as opposed to the latter, the optical $k$-vectors do not refract, and it is beam trajectories that are illustrated by the blue lines in the figure. The outcome is independent of the longitudinal position of the cell, as seen by the resemblance of **c** and **d**. Note that this lens is limited to $|\mathbf{k}_\perp| \ll k_z, k_0$ and is therefore not applicable to the evanescent components of the image [30].

As a geometrical effect, paraxial diffraction is manifested via $k$-dependence of the dispersion curve. The optimal approach to eliminate it is to superimpose dispersion with the same form and the opposite sign that is uniform in real space. An EIT vapor medium with a buffer gas and a uniform pump manifests these properties. The pump provides the anisotropy needed to single out the transverse wave-vector, while Dicke narrowing accounts for the quadratic *k*-dependence. The strength, frequency, and orientation of the pump control the magnitude, sign, and direction of the resulting diffraction. The possibilities may be further extended by introducing a non-uniform pump. The phase gradient of the pump then acts as a vector potential for the probe, in an effective Schrödinger dynamics of a charged particle. Currently, inherent loss limits the effectiveness of the scheme, *e.g.*, in terms of resolution. Resolving this limitation by combining linear gain, probably by utilizing more elaborated Raman processes, would clear the way to a vast variety of applications, in microscopy, lithography, switching, and more.

**Acknowledgments**

We thank Dimitry Yankelev for assistance with the experiments and Yoav Erlich for technical support. We thank Rami Pugatch for helpful discussions and suggestions.